# The Impact of High-Resolution Soil Moisture States on Short-Term Numerical Weather Prediction of Convective Initiation over South Africa.

Edward H. Engelbrecht,[a,b] ,Willem A. Landman [b, c], Stephanie Landman [a]

[a] South African Weather Service*, Centurion, Gauteng, South Africa*

[b] University of Pretoria*, Pretoria, Gauteng, South Africa*

[c]International Research Institute for Climate and Society, *The Earth Institute of Columbia University, Palisades, New York*

*Corresponding author*: Edward H. Engelbrecht, edward.engelbrecht@weathersa.co.za

# ABSTRACT

The interaction between the Earth's surface and the atmosphere plays a key role in the initiation of cumulus convection. Over the land surface, a necessary boundary condition to consider for resolving land-atmosphere interactions is soil moisture. The aim in the study is twofold. One, through object oriented and traditional verification techniques determine how higher resolution soil moisture initial conditions influences the prediction of the location and timing of convective initiation (CI) within a convective permitting, operational NWP model over South Africa. Two, to study the modelled CI-soil moisture relationship during real afternoon thunderstorm events. The study reports the results from 66 Unified Model simulations (at 4.4km grid resolution) for nine summer afternoon CI events during synoptically benign conditions over South Africa. The higher resolution soil moisture conditions reduce centroid distance between observed and forecast storms on average by 7km (9% improvement), with the most decrease in centroid distance occurring at the shortest lead times, by 12km. Most improvement in location error occurs in the zonal directional. However, little to no difference is found in the timing of CI, most likely attributable to the dominant effect of model grid size on CI timing, overshadowing the influence from soil moisture anomalies. Probability of CI is highest over dry and moderate soils and areas along distinct soil moisture gradients. The conclusion is that modelled CI over South Africa preferentially occurs on the periphery of wet soil moisture patches, where there is increased surface convergence of wind and higher sensible heat flux.

# 1. Introduction.

Fluxes of heat and moisture at the Earth's surface play a critical role in many meteorological phenomena that arise due to earth-atmosphere interactions (e.g Pielke, 2001; Stensrud, 2007). One such phenomenon is convection, where studies have shown that small-scale atmospheric boundaries play a key role in the location and timing of convective initiation (CI) (Maddox *et al*. 1980; Ziegler and Rasmussen 1998). The development of these small-scale atmospheric boundaries depends on changes in land surface properties (McCumber and Pielke 1981; Trier *et al.* 2004). Distinct heterogeneities within land surface properties can result in the development of mesoscale atmospheric circulations that initiate convection similar to the circulations present within land-sea breezes (Segal and Arritt, 1992; Dirmeyer and Shukla, 1993; Dirmeyer, 1995).

One variable that has been shown to create these land surface patterns necessary for the circulations to develop is soil moisture. Soil moisture is key to understanding land-atmosphere interactions due to its ability to partition incoming radiation from the sun into sensible heat flux for surface temperature increase and latent heat flux for evapotranspiration (Pielke, 2001). Eltahir and Pal (1996) hypothesized that sensible and latent heat flux are the pathways through which soil moisture may regulate atmospheric variables such as humidity, radiation and temperature of the planetary boundary layer (PBL) and thus its role in the dynamics and triggering of convection. Latent heat flux from increased soil moisture (as a result of rainfall) supplies the lower atmosphere with water vapour, increasing moist static energy in the PBL. This destabilizes the lower atmosphere further, resulting in favourable conditions for convection to initiate, which produces rainfall and thus resulting in a positive soil moisture-precipitation (SMP) feedback.

*a. Soil moisture-precipitation feedbacks*

SMP feedbacks are known to be of particular importance over semi-arid regions (Koster *et al.,* 2004; Dirmeyer, 2011; Taylor *et al.,* 2007; Taylor *et al.,* 2011), and have been found to be applicable to land-atmosphere interactions at a broad range of scales; from the continental climate (New *et al.,* 2003; Findell and Eltahir 2003a; Schlemmer *et al.,* 2011) down to the convective scale (Taylor *et al.,* 2011a; Barthlott and Kalthoff 2011; Froidevaux *et al.,* 2014). However, at the convective scale SMP feedbacks present a significant challenge to numerical weather prediction (NWP), especially concerning the prediction of thunderstorms that develop over uniform terrain where soil moisture accounts for the majority of surface variability (Froidevaux *et al.,* 2014).

*b. Motivation*

Accurate prediction of convection over South Africa is of significant importance due to the large number of intense thunderstorms that the country experiences (e.g Zipster *et al*., 2006; Gijben *et al*, 2017), often resulting in significant damage to infrastructure, agriculture, transportation and loss of life. An important aspect to accurately predict severe thunderstorms is the consideration of location and timing of CI (Case *et al*., 2011).

Tang *et al., (*2013), highlighted that convective permitting models (CPMs) show considerable promise particularly when the location of CI was reliant on details of surface forcing (such as land/sea contrast, orography etc.), however the initiation of convection remains a scientifically challenging problem (Markowski and Richardson, 2010). It is therefore crucial to accurately represent fine scale moisture boundaries that may lead to the initiation of deep convection, and in doing so make full use of a CPM's potential. The effect of soil moisture on CI within a CPM over South Africa at short-term forecasts have been largely neglected, evident from studies mainly focusing on its effect at seasonal and climate scales using coarse resolution models (e.g. Cook *et al*., 2006; Mackellar, 2007).

The hypothesis in this study is that soil moisture affects deep moist convection over South Africa to the extent that using high resolution soil moisture states as initial conditions in a CPM will improve the short-term prediction of CI. In testing this hypothesis, the study also investigates the nature of the modelled CI-soil moisture relationship. The evaluation results may therefore be understood in light of the model's ability to simulate known CI-soil moisture interactions.

## 2. Data and Methodology

The study follows an experimental design by running two identical configurations of the Unified Model (UM) at 4.4km horizontal grid resolution in parallel over the South African domain (Fig. 4), differing only in the input of initial soil moisture states.

*a. Model description*

The UM is a non-hydrostatic numerical weather prediction model developed at the UK Met Office (Davies *et al*., 2005). The global analysis is provided four times daily by the Met Office (at 00, 06, 12 and 18 UTC) using the Global Atmosphere (GA) version 6.1 science configuration (Walters *et al.,* 2017). A regional configuration of the UM at 4.4km spatial resolution is used operationally at the South Africa Weather Service (SAWS) and is one-way nested in the GA simulation for a domain covering all of southern Africa (Fig. 1 left).

UM only partially resolves convection; the mass flux of the convective parameterization is limited to favour deep or strong convection to take place explicitly, but parameterizes shallow convection (Roberts, 2003). This hybrid convection scheme is used in an attempt to alleviate the problems caused by the two extremes of having either no convective parameterization or full convective parameterization.

*b. Soil moisture initialization*

The operational land surface model (LSM) used in the Met Office UM is the Joint UK Land Environment Simulator (JULES) and provides the lower boundary conditions to the UM. This LSM makes use of the Met Office Surface Exchanges Scheme (MOSES 2; Essery et al., 2003) to simulate hydrological processes on the boundary layer and subterranean.

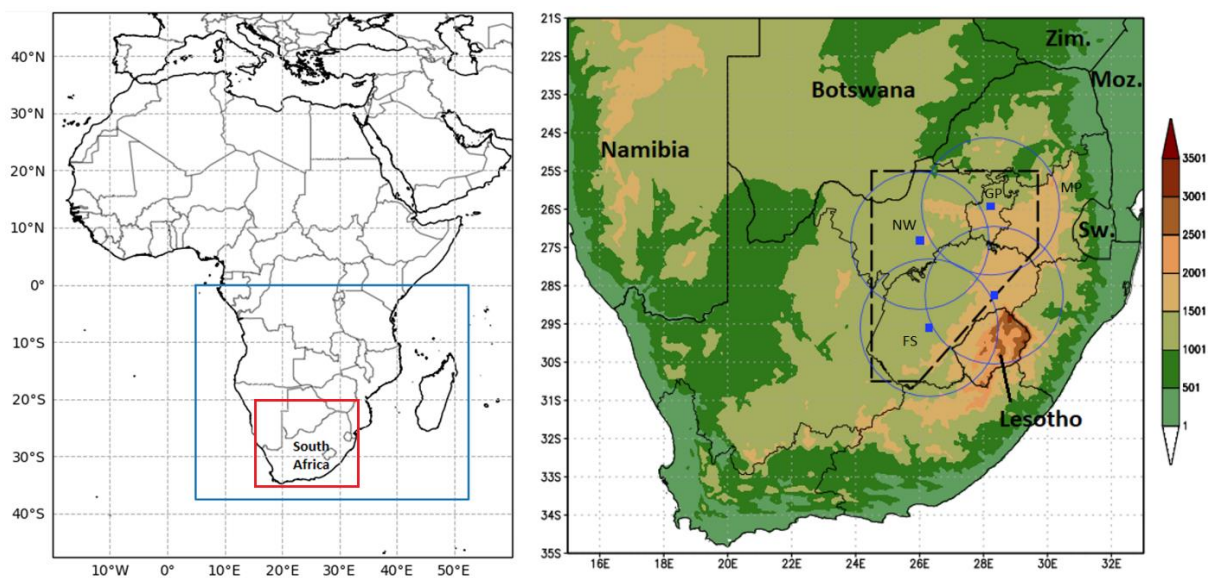

FIG. 1 **Left**: Location of the regional UM domain (blue square) as it is run operationally by the South African Weather Service (SAWS). **Right**: Verification domain depicted by the black dashed line. Blue squares mark the position of the radars while the blue circles depict their range of 200km. Zim. = Zimbabwe, Moz. = Mozambique and Sw. = Swaziland). FS. = Free State, NW. = North West, GP. = Gauteng, MP. = Mpumalanga. Shaded colours indicate height above sea level (m) of the orography.

Within JULES choices can be made (e.g. canopy radiation model type) and certain modules (e.g. vegetation dynamics) can be switched on or off to operate at different levels of complexity (Iwema *et al.,* 2016). This selection includes an option which allows the user to run the UM with JULES providing the soil moisture states at a horizontal resolution corresponding to the model's grid (without the need of dynamical downscaling). Furthermore, with this option switched on, the initial soil moisture conditions provided by the UM global analysis are replaced by soil wetness observations retrieved remotely via a passive microwave remote sensor (Advanced Scatterometer (ASCAT)) on board a satellite and assimilated with the modelled soil moisture. This is the soil moisture data set used in the experimental (referred to as "SM" from here on) simulations.

The control (CTRL) simulations for this study comprises of forecasts where this option is switched off (as is currently done at SAWS for operational use), which means that the soil moisture analysis is used directly from the lateral boundary conditions provided by the global analysis, at a horizontal resolution of ~10km.

In the UM, hydraulic relationships from van Genuchten (1980) are used to represent heat and water fluxes using a four-layer scheme. The thicknesses of these soil layers from the top down are 0.1, 0.25, 0.65 and 2.0m. For this study, only the top soil layer (0.1m) is used in the analysis. Based on soil moisture data for each event, each CI location point is assigned a soil moisture data grid cell using a nearest-neighborhood technique (Fig. 2a).

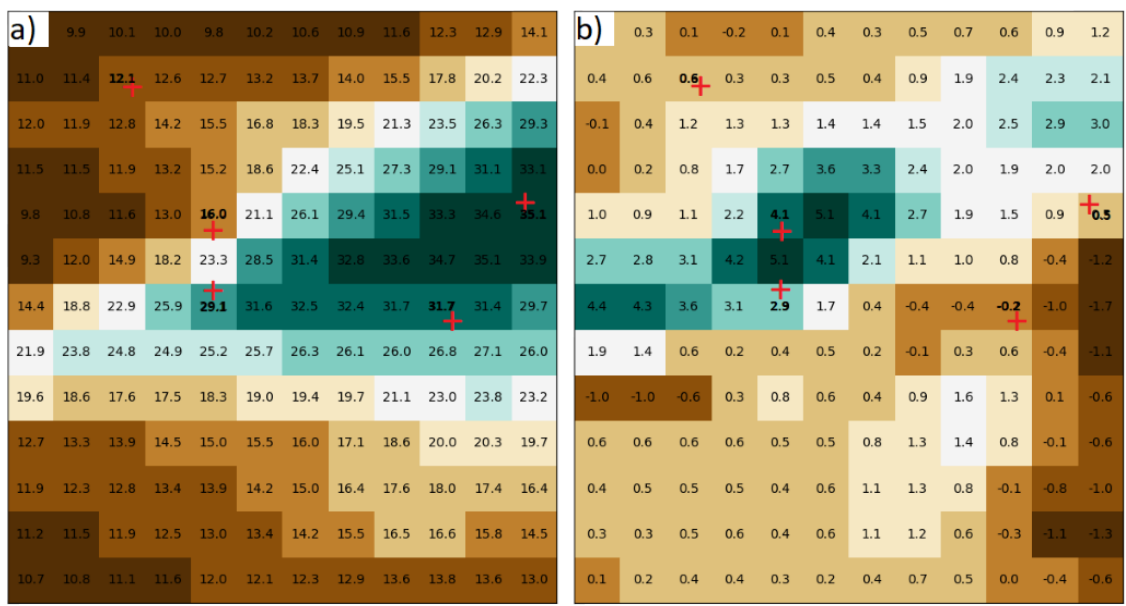

FIG. 2: Example of convection initiation (CI) location points (red cross) at 1200 UTC 15 December 2018. The numbers represent (a) moisture content of soil layer (kg m$^{-2}$) and (b) moisture content of soil gradient (kg m$^{-2}$ km$^{-1}$) with bold numbers denote values associated with a CI point.

Soil moisture gradients are then calculated at each of these grid cells, using a centered difference method (Chapra and Canale 2002). The number of grid points to differentiate over are experimented with to investigate at which scale the strongest soil moisture gradients, associated with CI locations, are found. Differencing is done over 1, 2, 3 and 4 grid points, corresponding to ~4, 8, 12 and 16km distances respectively. That is, if S represents a value of soil moisture content, then differencing was performed at $S_{j-n}$ and $S_{j+n}$, where n of 1, 2, 3 and 4 were used (Fig. 3). Soil moisture and soil moisture gradient values are then assigned to the closest occurring CI point. The data is binned into equal intervals of 5 $kg/m^2$ groups (0-5 through to 40-45). Similar binning of each CI location is assigned a soil moisture and soil moisture gradient value based on the nearest grid cell. The values are binned into 5 $kg/m^2$ value groups (i.e 0-5, 6-10 through to 40-45). Similar binning of data is done to investigate CI for soil moisture gradient relationships (i.e. bin intervals of 0.5 $kg/m^2/km$).

Each CI location and its associated soil moisture and soil moisture gradients are then used to determine the role that soil moisture and soil moisture heterogeneities has on the development of deep convection within the UM model.

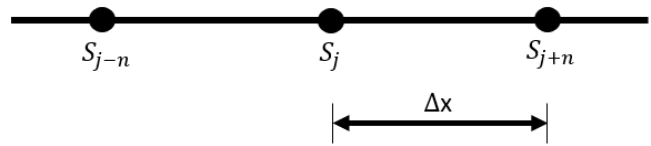

FIG. 3: Schematic of the grid points across which soil moisture gradients are calculated, where S represents the soil moisture content, j the latitude or longitude, n the number of grid points across which differencing is performed, and Δx the distance between grid points.

*c. Evaluation domain and event selection*

The study evaluates CI events occurring over a domain covering the central interior region of South Africa (Fig 1 right). Selection of the geographical region is not arbitrary; the chief motivation for pursuing this study is based on previous studies which have identified areas of transition zones between wet and dry climates as "hot-spots" for soil moisture-precipitation coupling (Koster *et al.,* 2004), where latent heat fluxes are most sensitive to variations in surface layer soil moisture (Dirmeyer, 2011). One such "hot-spot" occurs over southern Africa. However, due to a lack of radar coverage in the rest of southern Africa, the area is confined to where this "hot-spot" is located within South Africa. To minimize the effects of topography on the analysis of soil moisture-convection interactions, the most prominent terrain gradients along the eastern and south eastern parts of South Africa are excluded from the domain

CI events are chosen that reflect "weakly forced thunderstorms". Miller and Mote (2016) characterizes the environments favorable for this type of convection as an airmass lacking any lifting mechanisms on the synoptic scale. As a result, thunderstorms occurring under these conditions are mostly fostered by the combined effects of strong and subtle variations in surface temperatures, wind and moisture. Under these circumstances, the impact of the interaction mechanism between soil and atmosphere on near-surface variables and the PBL structure become greater (Barthlott and Kalthoff, 2011), which should allow for more representative testing of the hypothesis. A total of nine synoptically benign CI events are selected during the 2019 – 2020 summer rainfall season. For each event the UM is run four times at the main synoptic hours (00, 06, 12 and 18 UTC) for the respective CTRL and SM configurations. This gives a total of eight model runs for each event: four for each of the CTRL and SM configurations. In total 66 model runs were conducted for the 9 events. The lead times range from short to very short; the 12 and 18 UTC cycles are used to predict the CI event of the next day, which has lead times of about 24 to 32 hours. The 00 and 06 UTC model run predict the event on the same day as the model cycle, thus with lead times of about 6 to 12 hours. The ensemble of forecasts for each event can therefore be thought of as a lagged ensemble.

*d. Verification*

For each event a list of observed and modelled CI objects is created from the CTRL and SM sets of forecasts. Each set of forecasts are compared to radar observations and verification statistics are calculated using the Method for Object-based Diagnostic Evaluation (MODE) software (Davis *et al.,* (2006a). Verification of the location comprises of two components, namely 1) centroid error using MODE software and 2) zonal and meridional component of the location error. Because focus is placed on the initiation of convection, verification is performed only on the first storms that occur on the given day (similar to e.g. Duda and Gallus, 2013).

1) DEFINING CI

There is no generally accepted definition of CI relative to a specific radar threshold. Definitions of the instance when convection initiates range from a column maximum threshold of 30 (Wilson and Schreiber, 1986; Wilson and Mueler, 1993) to 35 dBZ (Mecikalski and Bedka, 2006). Various other approaches have since been introduced. Kain *et al*., (2008) used radar reflectivity of ≥ 35 dBZ at the -10°C isotherm level. A more recent study over South

Africa defined deep convection as the point where the 10dBZ value reached a height of 8 km (Keat *et al.,* 2018). In this study we use a threshold of ≥ 35 dBZ to define CI. This follows a similar approach by previous studies investigating similar phenomena (e.g. Duda and Gallus, 2013; Burghardt *et al.*, 2014).

2) OBJECT MATCHING

All thunderstorms occurring within a 150km radius (between the forecast and observed fields) are considered in the matching process. MODE then calculates pair attributes from the forecast and observed objects within this radius, which is used to determine the total interest between the forecast and observed counterparts. The total interest is summarized into a total interest value, between 0 and 1. The attributes that contribute to the interest value between the forecast and observed objects are user specified as shown in table 1. Storm intensity is not considered a key feature in this study, since we are mainly concerned with location error, but is still important to match thunderstorms with one another that are within a similar life cycle. Similar reasoning is applied when considering a weight for the areal size of storms. Therefore, storm intensity and areal size are given lower weights. However, centroid distance (C error) between the radar observed and modelled CI object, is given the most weight, so that preference is given to thunderstorm proximity. Only forecast and observed objects with an interest value of ≥ 0.6 are then considered a matched pair, or a hit.

TABLE 1: MODE fuzzy logic weights assigned to pair attributes to calculate the "total interest" value.

| Object attribute | Weight (%) |
|---|---|
| Centroid distance | 50 |
| Intensity | 20 |
| Areal coverage / object size | 30 |

## 3. Results

Although not shown, we also investigated the impact of higher resolution soil moisture initialization data on the timing of modelled CI but found on average little to no difference to the delay in CI compared to the CTRL simulations. This is most likely attributable to the dominant effect of model grid size on CI timing, overshadowing the influence from soil moisture anomalies

*a. Impact on location of CI*

For each model run, the C error for all matched storms are averaged at each 6-minute forecast intervals, starting at the time of CI and ending one hour after initiation of CI. To visually demonstrate the spread and shape of the C error distribution, the average C error at each time step is collected into 20km bins and their frequency counted (Fig. 4). This distribution is comprised from a total of 215 and 209 matched storms for the CTRL and SM simulations respectively.

Event C error ranges from 45 to 138 km with an average of 82 and 76 km for the CTRL and SM forecasts respectively. The results here are similar to C errors found by past studies such as Burghardt *et al* (2014) and Duda & Gallus (2013) where an average storm C error of approximately 48km was found in the former study and 105km in the latter. The 12 and 06Z model cycles had the highest frequency of outliers for the 140-150km bin, and they also share a similar frequency distribution pattern.

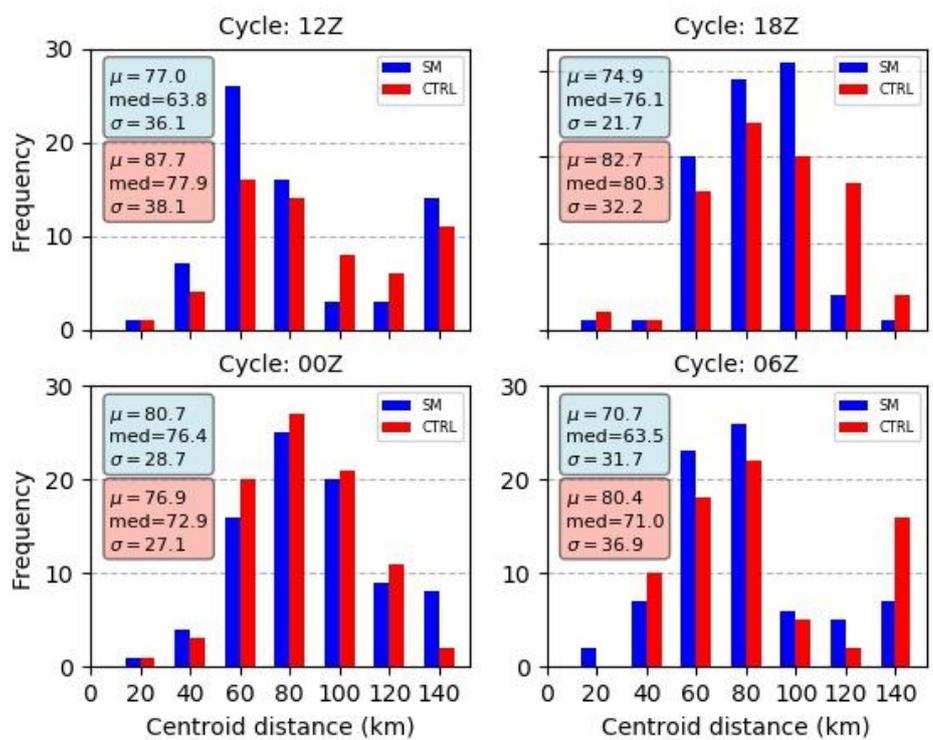

FIG. 4: Frequency histograms of C error for matched CI objects in each model cycle of the SM and CTRL predictions. The data is based on 33 model runs for 9 convective events. μ = mean, σ = standard deviation.

The SM simulations improves on the C error by 11km for the 12 UTC model cycle, 8km for the 18 UTC model cycle, -4km for the 00 UTC model cycle and 10km for the 06 UTC model cycle. Although not a very large improvement, the improvement in C error is consistent (apart from the 00UTC run) throughout all model cycles. The 00 UTC results however are not significant, whereas slight to high significance levels are associated with the other cycles.

A possible reason for the lack of significance between the CTRL and SM C error of the 00 UTC cycle could be explained by the more comprehensive global observation data that the initial conditions for the 00 UTC run is comprised of. Thus, owing to the already comprehensive observations, the higher resolution soil moisture may make little difference, where-as with more data sparse model cycles such as the 06 UTC run, the improvement from the SM simulations is more noticeable. Furthermore, the shape of the C error distribution in both the CTRL and SM simulations for the 00UTC cycle is closest to a normal distribution compared to the other cycles, with fewer outliers and a smaller standard deviation.

The results in figure 4 for the 12 UTC cycle is only significant at the 85% level, increasing from 95% for the 18 UTC cycle and 90% for the 06 UTC cycle, which suggests that the impact of the improved soil moisture states are influenced by how close in time the model cycle is to the convection event. In order to investigate the impact of the SM simulations on the respective zonal and meridional components of the distance errors, a kernel density estimate (KDE) is fit to the collection of average storm distance errors for each model run, separated into the CTRL (Fig. 5a) and SM (Fig. 5b) simulations. These include model runs of all cycles (12, 18, 00 and 06 UTC); 33 model runs for each of the CTRL and SM simulations. The primary motivation for using a KDE fit is to aid in visualizing the distribution of each model run's CI distance errors and accentuate the zonal and meridional component of the errors.

The CI locations here include matched and unmatched storms, which is why there are several outliers. The mean absolute error for the CTRL and SM runs is 79 and 73km respectively. Comparing the CTRL and SM run's KDE distributions it is noticeable that the distance errors for the individual model runs are more densely clustered closer to the $y = 0$ line in the SM simulations and only slightly closer to the $x = 0$ line, compared to the CTRL simulation where the highest frequency occurs slightly further to the north.

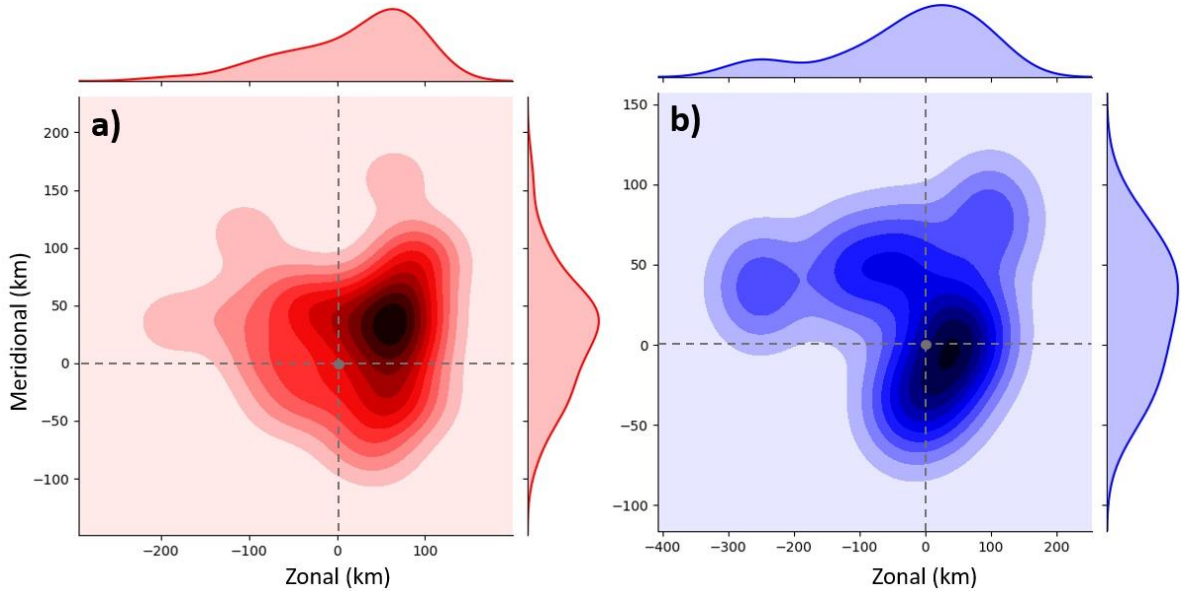

FIG. 5: KDE distribution of unmatched CI location error (km) for each model run. a) CTRL and b) SM simulations.

*b. Soil Moisture-CI relationship*

To give further context to these results, the modelled relationship between soil moisture and CI location is investigated. That is, the extent to which the UM at 4.4 km horizontal resolution is able to simulate known soil moisture CI relationships is examined. In addition, this investigation will serve to answer the question as to whether the amount of improvement in C error is to be expected given the nature of the modelled relationship between soil moisture and CI found here.

This part of the study follows a similar methodology to work done by Frye and Mote (2010) over the Southern Great Plains region in North America. The main difference being that the relationship between soil moisture and convection investigated here is a modelled relationship, whereas in Frye and Mote (2010) it is observed. We also investigate how the relationship differs between various spatial scales.

The CI points for all model runs of the SM simulation is plotted in figure 6. There appears to be a mostly uniform (random) spatial distribution of initiation points across the domain, with limited clustering. The exception is the far eastern parts of the domain where there is increased clustering. This clustering is likely due to the eastern boundary being too close to the escarpment/mountain range there, and therefore its influence triggers CI disproportionately in that region. This phenomenon is a good example for the motivation behind why in the methodology we aimed to exclude as much of the mountainous areas as possible. As a result of this clustering the domain under examination is shifted to the 30°E longitude to exclude the dense CI clustering. The other exception to the uniformity is the southern, and far north central parts of the domain where there is a definite area clear of modelled CI.

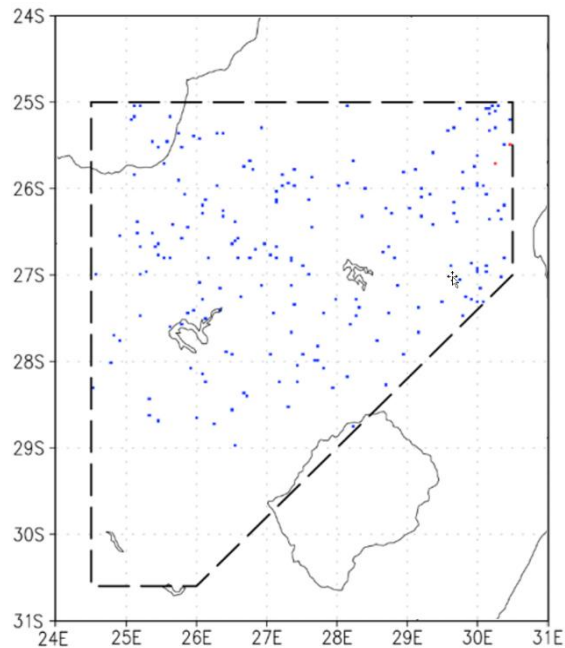

FIG.6: All CI location points within the evaluation domain for the SM simulation across all model cycles.

The total number of CI points is 285 for the 33 SM model runs. This number is similar to the amount of CI points found in the research of Frye and Mote (2010), where they found 359 storms over 43 synoptically benign days. As is discussed in more detail in section 2.b, the soil moisture values associated with the initiation points in figure 6, based on the grid cell they are nearest to, are grouped into equal interval bins of 5 kg m$^{-2}$ (table 2). Average soil moisture across the domain between the different model cycles differs very little, ranging between 9.7 to 10.7 kg m$^{-2}$.

Because of the non-normal distribution of soil moisture values in table 2, as was also found in Frye and Mote (2010) where it decreases linearly as soil moisture content increases, the probability of convection to occur within each soil moisture bin is calculated rather than the total occurrences of the soil moisture values (Fig. 7).

Average probability for convection to occur within any given cell box is 0.056%. The probability appears very small; however, this is due to the small spatial grid of the model, resulting in a large number of grid points/soil moisture values in relation to the number of CI points. What is important is the shape of the graph of figure 7, which plots the probability of CI across the range of soil moisture value bins.

TABLE 2: Number of occurrences of all Soil Moisture Content values (kg m$^{-2}$)(ALL) and those points associated with CI.

| Bins | ALL | CI |
|---|---|---|
| 5 | 263508 | 62 |
| 10 | 263198 | 67 |
| 15 | 215044 | 72 |
| 20 | 165314 | 49 |
| 25 | 76658 | 25 |
| 30 | 14064 | 9 |
| 35 | 2248 | 1 |
| 40 | 268 | 0 |
| 45 | 14 | 0 |

The graph shows probability of convection to be relatively constant between 0.02 - 0.03% up until a soil moisture value of 25 is reached, where after it increases rapidly up to 0.07% at a value of 30 kg m$^{-2}$. Thereafter, as soil moisture increases further, the probability of CI drops to 0% in the wettest spectrum of soil moisture values.

Similar to table 2, table 3 shows the occurrence of all soil moisture content gradient values across the domain, binned into equal intervals of 1 kg m$^{-2}$ km$^{-1}$, as well as gradient values associated with CI, calculated across different gradient distances. Again, similar to the soil moisture content values, as the soil moisture gradient values increase, the frequency of the soil moisture gradients decreases. That is, the frequency of gradient values is greater at smaller values. Hence, the percentage of CI over each soil moisture gradient bin is calculated, thereby showing the amount of CI in proportion to the amount of gradient values across the domain (Fig. 8).

Figure 8 illustrates the probability shape of CI across the spectrum of soil moisture gradient bins. This shape differs markedly from the pattern in figure 7, where instead of the probability of CI decreasing at some point as soil moisture increases, the probability (after decreasing at medium gradients) increases notably at higher soil moisture gradient values.

This shape remains similar across gradients from 4 to 16 km, indicated by the average probability of CI across all gradients (black dashed line), though the relationship appears strongest at the 4 and 8 km gradients.

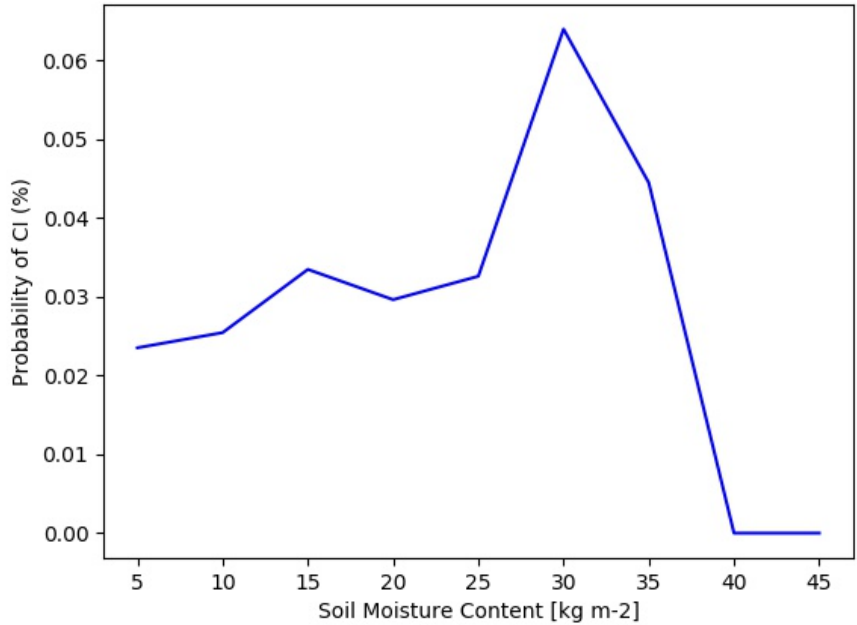

FIG. 7: Probability of convection initiation by Soil Moisture Content (kg m$^{-2}$). Probabilities are filtered into intervals of 5 kg m$^{-2}$ bins.

To briefly demonstrate how the (modelled) atmospheric response to soil moisture heterogeneities lead to the patterns in figures 7 and 8, and how it may influence CI location, an example from the 15 December 2018 CI event is investigated, using the 00 UTC, SM simulation, cycle run. Figure 9 illustrates the distribution soil moisture content at T+9 over a small area of the evaluation domain, three hours prior to the initiation of a deep, moist convective cell located at the green marker.

Over the central parts of figure 9 there is a distinct wet soil moisture patch, roughly 60km across, surrounded by drier soils. Along the edges of the domain, particularly the top left corner (25.7S, 27.2E), there is another, though much smaller, wet soil moisture patch.

Visually comparing the blue contours with soil moisture shows a clear relationship between surface wind divergence and wetter soil patches, such that wetter soils are marked by diverging surface winds (positive values) and drier soils associated with convergence.

TABLE 3: Number of occurrences of all Soil Moisture Content values (kg/m$^2$)(ALL) and those points associated with CI. 4, 8 12 and 16km represents the distance of the gradients performed.

|     | ALL/4km | ALL/8km | ALL/12km | ALL/16km | CI/4km | CI/8km | CI/12km | CI/16km |
|-----|---------|---------|----------|----------|--------|--------|---------|---------|
| 1.0 | 753770  | 556014  | 435220   | 354490   | 191    | 134    | 103     | 80      |
| 2.0 | 181273  | 298953  | 356244   | 372950   | 72     | 102    | 110     | 100     |
| 3.0 | 44410   | 94449   | 133637   | 171662   | 16     | 34     | 52      | 67      |
| 4.0 | 13408   | 29882   | 45628    | 61137    | 5      | 9      | 13      | 22      |
| 5.0 | 4219    | 11201   | 16109    | 22997    | 2      | 1      | 2       | 8       |
| 6.0 | 1622    | 4535    | 6627     | 8934     | 0      | 5      | 4       | 8       |

The primary reason for this result is further illustrated in Figure 10a, showing a high Bowen ratio (~2.6) over the drier soils where the sensible heat flux is high (400 $Wm^{-2}$), and the latent heat flux considerably lower (150 $Wm^{-2}$).

Due to the increased surface heating (relative to the area over the wet patch at 28 °E), air rises disproportionately more over the dry soils than it does over the wet soil, seen in the vertical velocity field (Fig. 10b). Horizontal surface winds compensate by replacing the thermals over the dry soils, thereby creating the wind (and divergence) pattern seen in figure 10b.

The response of the PBL is further illustrated in figure 11b, where the PBL height is at 2800m over the dry soils, but closer to the wet patch towards the east, the latent heat flux increases while the sensible heat flux decreases, resulting in more stable air (Fig. 10b) due to lack of vertical motion and a thus a lower PBL height. These are very similar circulation patterns seen in land-sea breeze effects, because it is the same mechanisms that govern both, i.e. differential surface heating.

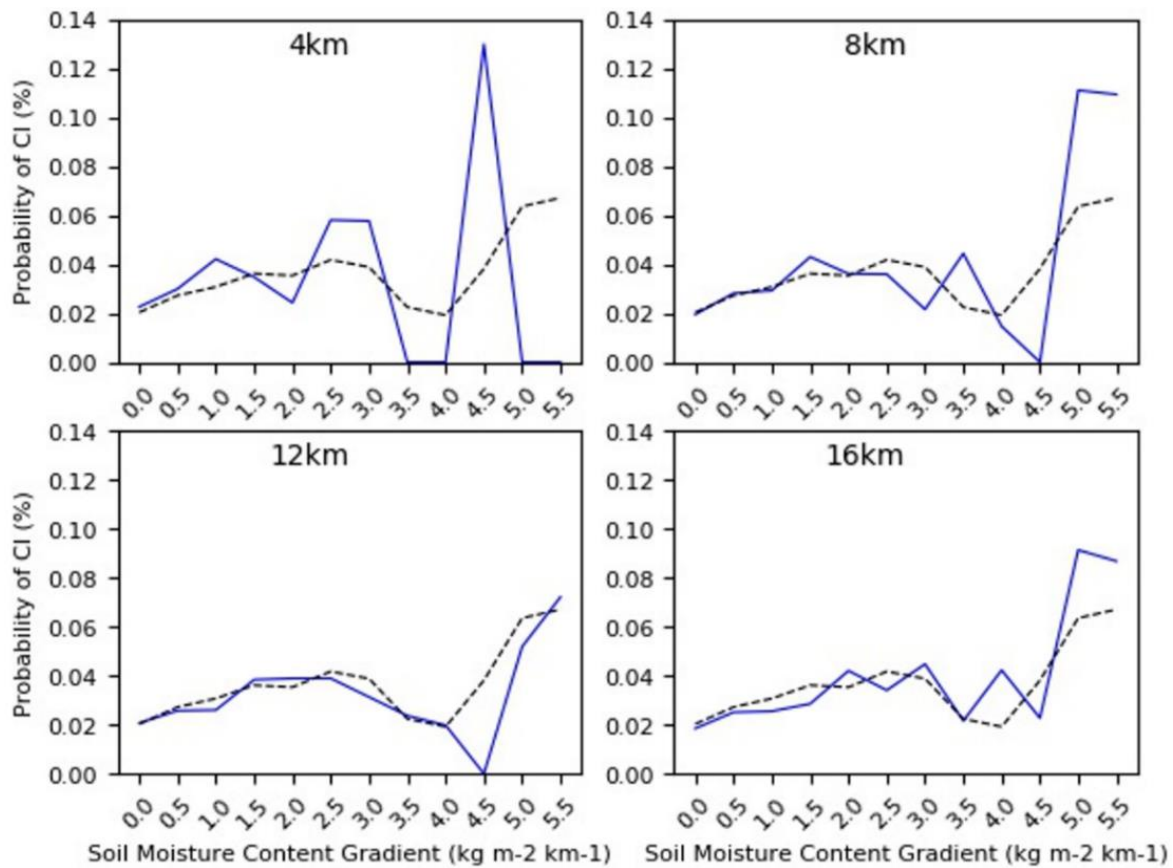

FIG. 8: Probability of convection initiation by Soil Moisture Content gradient (kg m$^{-2}$km$^{-2}$). Dashed line denotes the average probability of CI (%) across all the different gradient distances.

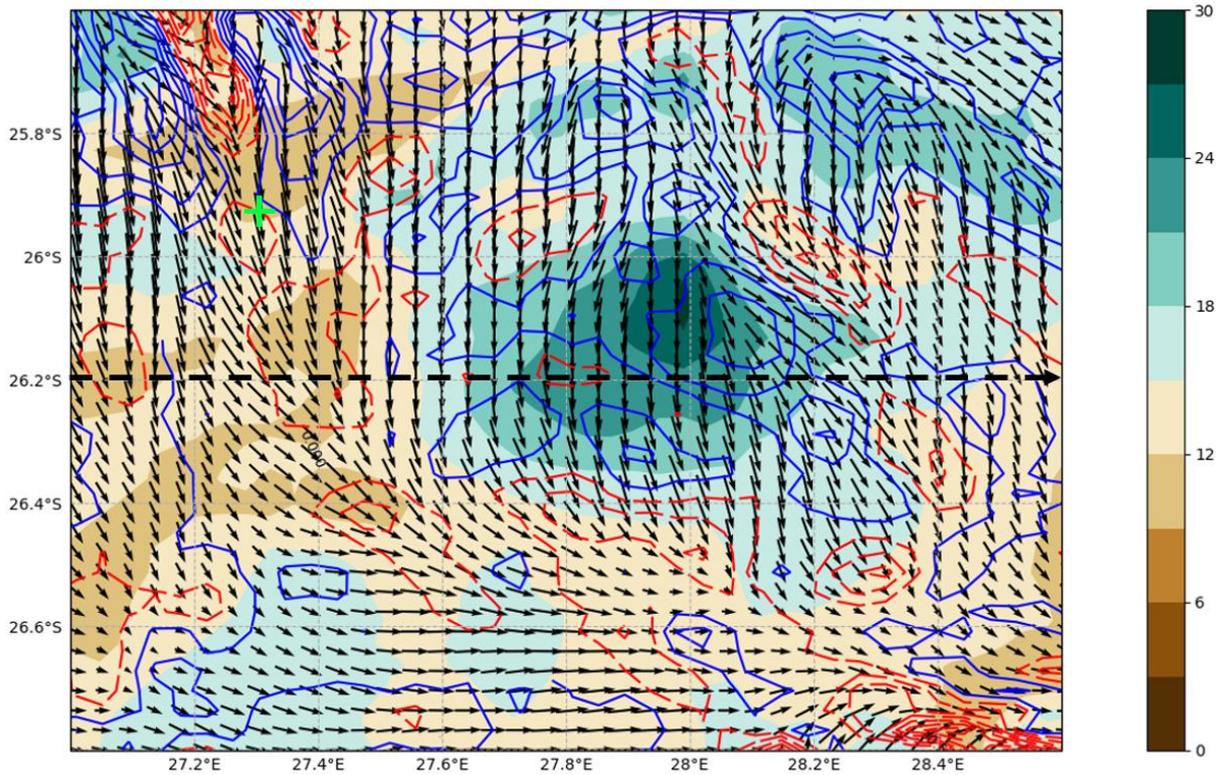

FIG 9: Soil moisture content (shaded; kgm$^{-2}$) prediction valid for 09 UTC, 15 December 2018 with 10-m surface wind prediction valid for the same time represented by vectors. Blue (red) contours represent surface wind divergence (convergence). The black dashed line indicates the latitude used for the vertical profile in figure 10. Green marker indicates the location of a CI point.

## 4. Discussion

The lack of impact on CI timing compares similarly to findings from a soil moisture sensitivity study on rainfall by Barthlott and Kalthoff (2011). The authors found that the first occurrence of rainfall between model runs with varied volumetric soil moisture initial conditions occurred between 0900 and 1000 UTC, regardless of the wetness of the initial soil conditions (by perturbing it between -50% and +50%). Thus, not a significant difference either. A reason for this could be that the effect of grid spacing in a CPM plays a dominant role in the timing of convective initiation.

The results from the impact of soil moisture on the C error show that the closer the model cycle is to the convection event, the more noticeable the influence of the improved soil moisture states will become. If this is true, then an explanation for the increase in significance percentiles from 1200UTC to 0600UTC is that in the event of soil moisture-convection feedbacks becoming noticeably strong (such as under weak synoptic forcing), model rainfall from thunderstorms occurring in the afternoon/evening of the 12 and 18UTC cycles may in turn affect the model thunderstorm locations the next day. Thus, erroneous thunderstorm locations from the previous day can further initiate convection along boundaries of wet soils, thereby compounding the errors in CI location. Furthermore, the model rainfall will also contaminate the soil moisture states gained from recent observations (i.e satellite), thereby removing the advantage of the high-resolution satellite observations.

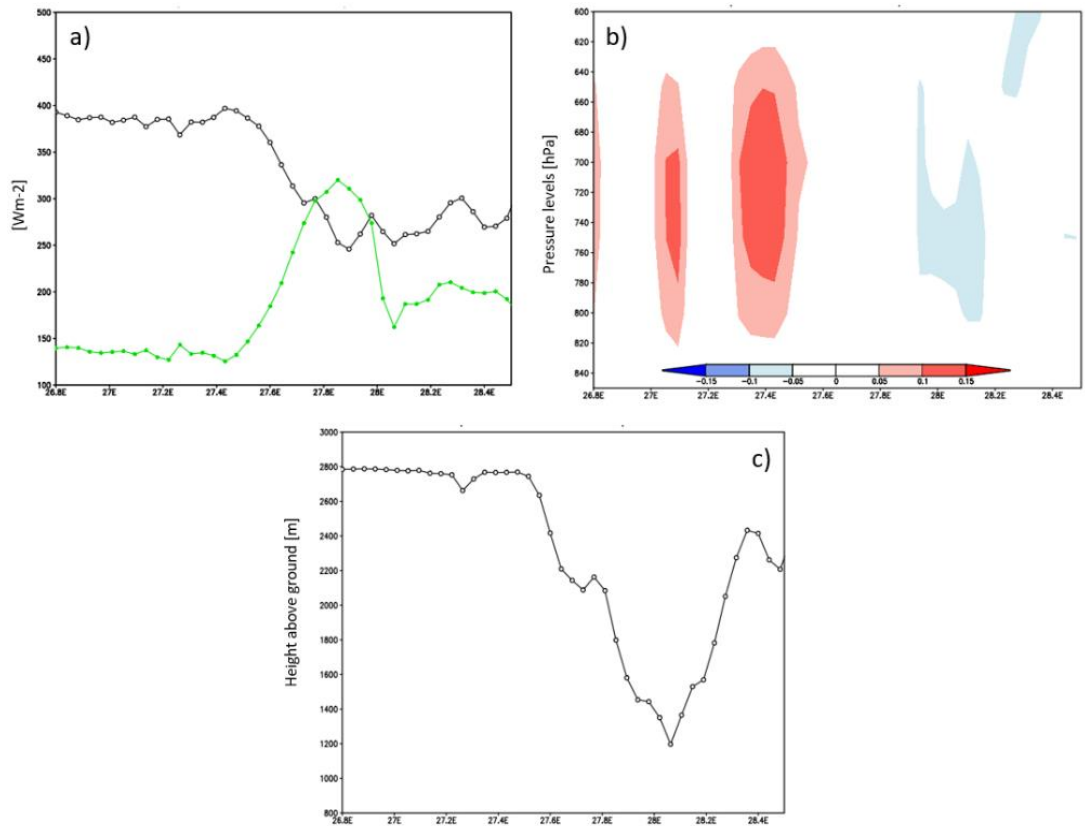

FIG. 10: a) Sensible heat flux (Wm$^{-2}$; black line) and latent heat flux (Wm$^{-2}$; green line), b) vertical velocity (m/s) where positive values indicate upward motion, c) height of PBL (m) above ground level.

An example of the model cycle differences in soil moisture can be seen in figure 12, showing the difference in top layer soil moisture between a 12 UTC model cycle run and 06 UTC model cycle run, both valid for the same time at 09 UTC on 22 February 2019, which is T+21 for the 12Z cycle run and T+3 for the 06Z run. Positive values indicate where soil moisture is higher in the 12 UTC run compared to the 06 UTC run. The differences shown here are due to the modelled thunderstorms at 12 UTC resulting in much heavier rainfall and over a larger surface area than what materialized and observed via satellite. This erroneous soil moisture will then further influence thunderstorm development at around T+24. This example also highlights the need to improve precipitation schemes in models, since the precipitation can feedback on itself when soil-moisture interactions are more pronounced.

When the meridional and zonal components of the distance error are analyzed, then it shows evidence that the higher resolution soil moisture conditions improved the zonal placement of CI more so than compared with the meridional direction. This result may be because on average (climatologically) the soil moisture gradient over South Africa is strongest in the longitudinal direction and where most variation takes place.

The graph pattern in figure 8 shows a distinct negative relationship between soil moisture values and CI. However, in the observations of Frye and Mote (2010) the relationship under synoptic benign conditions are less clear, showing no particular soil moisture values which favours CI. Their results are interesting since other observational studies, especially from Taylor *et al.,* (2011) and Taylor (2015), found distinct negative CI-soil moisture relationships (over the Sahel and Europe), where CI tended to favour drier soils. This highlights the ongoing uncertainty in the CI-soil moisture relationship "problem".

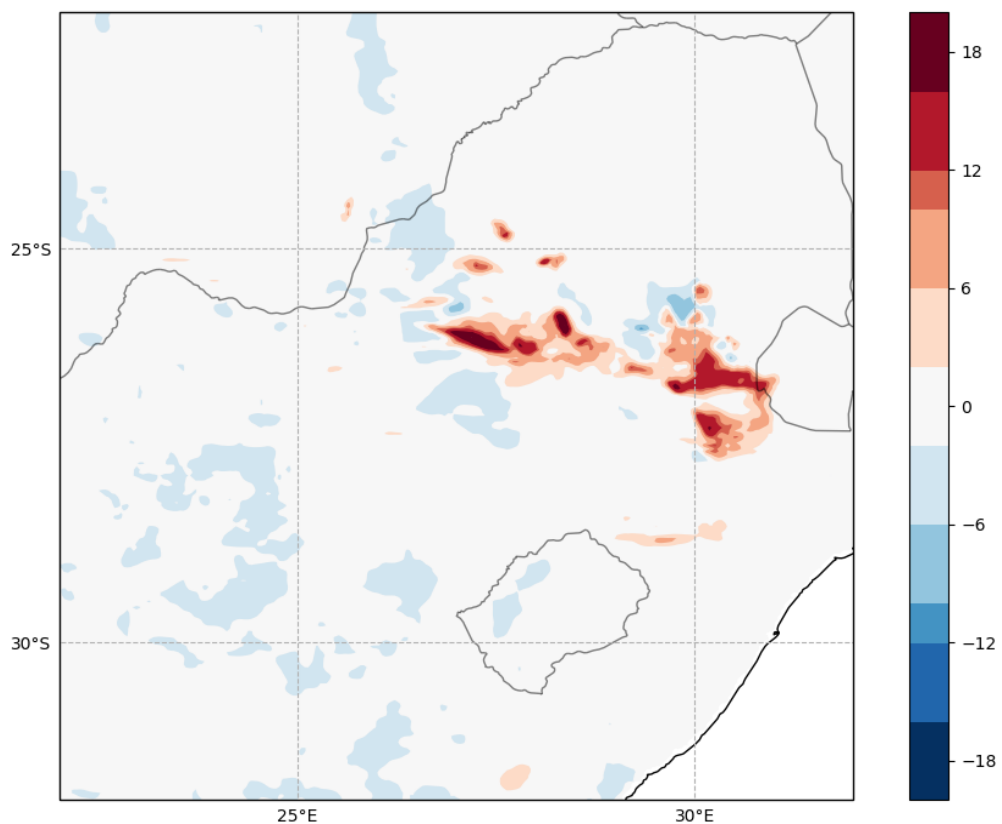

FIG. 11: Difference in top layer soil moisture content (kg/m$^2$) prediction between the 12 UTC 14 December 2018 and 06 UTC 15 December 2018 model runs, both valid for 09 UTC 15 December 2018 (12 UTC minus 06 UTC soil moisture prediction)

However, there can also be a number of reasons for the differences in the graphical patterns. For one, Frye and Mote (2010) speculated that the increased probability of CI over wet soils on synoptically benign days may be due to the lack of a low level jet (LLJ) to provide a moisture source, and as a result, in the absence of it the wetter soils provided the necessary moisture source to destabilize the atmosphere. Secondly, the model grid size may favour strong thermals for the initiation of convection (as opposed to being more sensitive to instability caused by moist static energy), to surface flux parameterizations. All of these possibilities are speculative but may provide insight to future investigation.

An important consideration to take into account is the fraction of CI, relative to the total number of CI, which are influenced by soil moisture gradients. Figure 10 shows that, although there is a positive linear relationship between soil moisture gradient and probability of CI (especially at 4km gradient length), it is only up to around 20 storms (out of 285; table 3) that are initiated along very high soil moisture gradients.

This result may further explain why there is only a relatively small improvement in the overall C error (of ~7km) by the higher resolution soil moisture initial states. However, it remains unclear whether the SM model runs with higher improvement are also associated with higher soil moisture variability across the area where thunderstorms occurred, and whether in those cases more CI are associated high soil moisture gradients.

Comparing the soil moisture gradient results (Fig. 8) to those from Frye and Mote (2010) shows a lot more similarities than for soil moisture alone. The reason why the pattern is similar within the gradients of soil moisture could be because the location (South Africa vs. North America) should not play a significant factor due to both areas marked by high frequency of deep moist convection during summer, and the fact that the relationship between heterogeneous soil moisture and convection should remain similar regardless of location, similar to land/sea breeze effects remaining constant regardless of location. The pattern in figure 8 is therefore strongly supported by observations evidence, and it can be concluded that the convective permitting UM at 4.4km horizontal resolution is successful at producing known CI-soil moisture relationships.

## 5. Conclusion

The interaction between the Earth's surface and the atmosphere plays a key role in the initiation of cumulus convection (Pielke, 2001). Over the land surface, a necessary boundary condition to consider in resolving land surface interactions is soil moisture. Numerous studies have regarded soil moisture second to sea surface temperatures (SST) in its ability to increase the predictability of the atmosphere (e.g. Dirmeyer 1995, Cheng and Cotton, 2002). These findings are largely due to soil moisture's control on the partitioning of solar radiation into sensible heat flux for surface heating and latent heat fluxes for evaporation (Pielke, 2001).

A large amount of research (both modelling and observational) has capitalized on improved soil moisture observations to investigate CI-soil moisture relations. However, in addition to the lack of such studies over South Africa, many of these studies which focused on the nature of the relationship have been either observationally driven or based on highly idealized modelling studies, with a lack of real events within an operational NWP model. The study reports the results from 66 model simulations for nine summer afternoon CI events during synoptic benign conditions over South Africa.

There are four key conclusions from the results:
- In most cases, the higher resolution soil moisture states are successful at improving the C error between forecast and observed CI by around 7km on average.
- Generally, C error from the SM forecasts decreases as the lead time to the CI event decreases.
- The 00 UTC cycle is the exception, where the SM forecasts increases the C error, albeit marginally and the difference between the CTRL and SM is not significant.
- The improved soil moisture states made little difference to the timing of simulated CI.

With regards to the soil moisture-CI relationship:
- CI occurs preferentially over dry to moderately wet soils.
- The probability of CI is inconsistent at low to moderate soil moisture gradients, but peaks over higher gradients at scales from 4 to 16km.

In addition to Frye and Mote (2010), the findings in this study regarding the soil moisture-CI relationship also agrees strongly with earlier notable studies such as and Cheng and Cotton (2002), Taylor *et al.,* (2011), Barthlott and Kalthoff (2011) and Froidevaux *et al.* (2014). That is, wet soils tend to suppress convection, but CI is focused along the peripheries of the wet patches where there is increased vertical motion, resulting in a negative feedback between wet soils and convective initiation.

The unique approach to test the mesoscale influence of soil moisture initialization on convective activity presented in this paper has taken NWP research in South Africa to a new level of operational application and use. Ultimately, the results show that efforts to improve the state of soil moisture initialization in regional NWP modelling over South Africa can lead to improved weather forecasts of thunderstorms over the region, which is often severe in nature. Further research over South Africa in this avenue can therefore be fruitful as spatial resolution of NWP models become increasingly higher, allowing for better representation of weather phenomena that result from land-atmosphere interactions.

*Acknowledgments.*

We would like to thank The Centre for High Performance Computing (CHPC) for providing access to their computing resources, free of charge, without which this study would not have been possible.